# Quantum circuit simulation with high-order optical tensor processing

Yufeng Zhang,[1]* Xuhan Guo,[2,3]* Zixuan Ning,[1] Zhiliang Yuan,[4] Yikai Su,[2,3] Kaizhi Wang,[2]* Sergei Turitsyn,[5] Zhipei Sun[1]*

[1.] Department of Electronics and Nanoengineering, Aalto University, Espoo, Finland.

[2.] School of Information Science and Electrical Engineering, Shanghai Jiao Tong University, Shanghai, China.

[3.] State Key Laboratory of Photonics and Communications, Shanghai Jiao Tong University, Shanghai, China.

[4.] Beijing Academy of Quantum Information Sciences, Beijing, China.

[5.] Aston Institute of Photonic Technologies, Aston University, Birmingham, UK.

Corresponding authors: Yufeng Zhang (yufeng.zhang@aalto.fi), Xuhan Guo (guoxuhan@sjtu.edu.cn), Kaizhi Wang (kz_wang@sjtu.edu.cn), Zhipei Sun (zhipei.sun@aalto.fi)

**Abstract:**

Classical simulation is essential for developing and testing quantum circuits, but the exponential growth of Hilbert space challenges conventional digital processors. Here, we introduce a classical optical quantum simulation architecture that treats gates and batches of input states as independent tensor dimensions rather than separate simulation instances, while mapping eligible tensor network contractions onto the same processor. Using co-designed multidimensional encoding and readout across spatial and wavelength degrees of freedom, we experimentally realize high-order tensor-matrix multiplication in a single propagation and process all 16 computational-basis inputs for each 4-qubit gate. Using a physics-based optical propagation model, we further numerically evaluate the architecture for the Deutsch algorithm and for tensor network of all 128 target instances of a 7-qubit Grover algorithm. These results establish a unified framework for gate-state batch and tensor network simulation, providing a route to exploit optical concurrency across quantum simulation tasks.

## Introduction

Simulation of quantum systems plays a central role in the development of quantum technologies, enabling the design and optimization of quantum algorithms and circuits(*1*), as well as the exploration of complex quantum many-body systems(*2, 3*). Despite rapid progress in recent years, practical quantum computers remain expensive, scarce, and constrained by stringent operating requirements(*4, 5*), limiting their accessibility to a broad community of researchers and developers(*6, 7*). Furthermore, a dedicated quantum system is unlikely to meet the needs of different quantum simulation tasks. As a result, classical simulation remains indispensable for quantum research at present. Digital simulation platforms have emerged as an interim solution by exploiting tensor formulations of quantum mechanics. Such platforms have enabled classical simulations of quantum circuits(*8-12*), quantum algorithms(*13*), and quantum machine learning(*14, 15*) through large-scale tensor processing. However, as the number of qubits or circuit depth increases, the associated computational cost and energy consumption scale exponentially. The underlying execution model of digital platforms remains fundamentally serial at the hardware level, rendering timely quantum simulation increasingly unsustainable, both economically and energetically.

In parallel, classical optical computing has emerged as a compelling classical alternative, offering intrinsically high bandwidth, ultrafast operation, and exceptionally low energy consumption(*16-18*). Crucially, optical systems naturally support multiple independent degrees of freedom, including space, momentum, wavelength, polarization, and topological encodings(*19*), providing a uniquely high-dimensional physical substrate for computation. These capabilities have enabled substantial advances in optical neural networks(*20-23*) and signal processing(*24, 25*), in some cases surpassing graphics processing unit (GPU)-based performance(*26-29*). A key direction in optical computing is the development of a general tensor computing paradigm, progressing from vector dot products(*30*) to matrix-vector(*31-35*), and matrix-matrix multiplication(*36-39*). Against this broader development, optical systems have long been explored for emulating quantum logic and algorithms, ranging from mode-encoded implementations of Grover and Deutsch algorithms using classical or quantum light(*40-43*) to recent demonstrations of the Deutsch-Jozsa algorithm through optical matrix-vector multiplication(*44*). Although these studies established the feasibility of optical quantum state emulation, they remained primarily low-parallelism and algorithm-specific, relying on mode-wise transformations or low-order vector operations. They therefore did not provide the arithmetic concurrency of optimized GPU matrix-matrix multiplication or a general framework for batched high-order tensor processing. Moreover, direct optical encoding of full quantum states retains exponential resource scaling with qubit numbers. Tensor network methods, particularly matrix product states (MPS), reduce this burden for structured and weakly entangled systems by relating complexity to bond dimension(*45*). However, their matrix-tensor contractions cannot be natively parallelized by existing optical matrix-vector processors. This raises a central question: can multiple optical degrees of freedom be jointly exploited for high-throughput, high-order tensor processing in quantum system simulation?

Here, we introduce a parallel classical optical quantum simulation architecture that treats gates and batches of input states as independent tensor dimensions and maps eligible tensor network contractions onto the same high-order optical processing (Fig. 1). We first introduce the operating principles and experimentally validate an optical processor capable of performing parallel tensor-matrix multiplications within a single optical propagation. We then experimentally demonstrate batched quantum state-gate operations across multiple wavelength channels, including gates acting on 2- to 4-qubit states. Using a physics-based numerical optical propagation model, we further extend the architecture to the Deutsch algorithm and to the tensor network contractions required for a MPS implementation of Grover algorithm and brickwork circuit. Systematic benchmarks over system size, bond dimension and circuit depth quantify the interplay between tensor network approximation and accumulated optical error. Together, these results establish high-order optical tensor processing for parallel gate-state batch and tensor network simulation as a unified architecture, enabling concurrent simulation across operator, state-batch and virtual-bond dimensions.

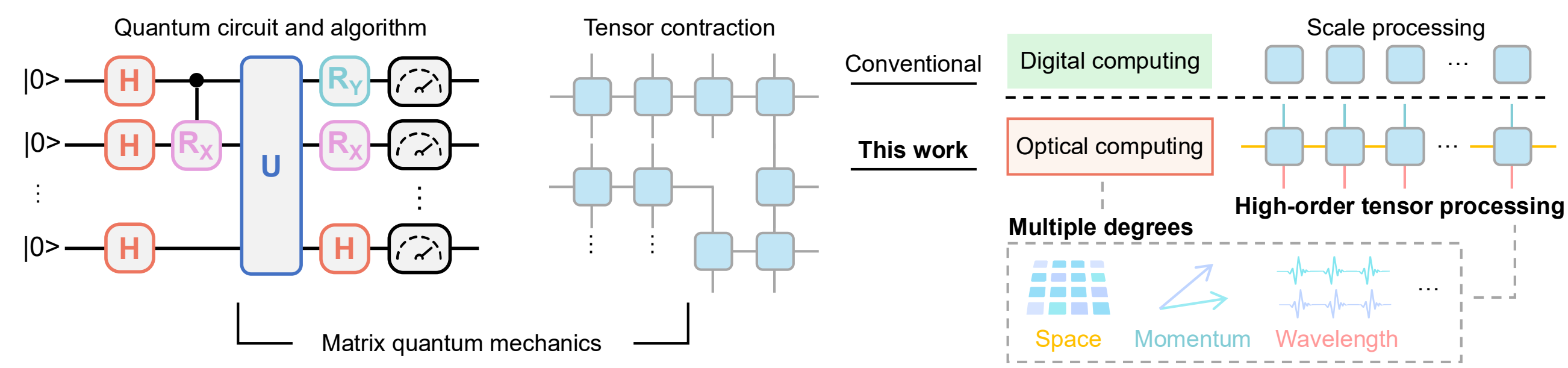

**Fig. 1 Concept of quantum circuit simulation with high-order optical tensor processing.** Digital quantum simulation typically involves modeling physical or computational problems to quantum circuits and algorithms, which are then executed through tensor contraction derived from matrix quantum mechanics, instead of real quantum hardware. In digital scalar computing, a tensor contraction needs many rounds of processing. In our optical architecture, spatial amplitude and phase encode the contraction dimensions, and wavelength supplies an independent operator dimension. Multiple gate-state instances are thereby organized as a high-order tensor processing whose optical evaluation occurs within a single propagation.

## Results

### Demonstration of high-order optical tensor processing

To establish the high-order computational primitive underlying the proposed gate-state batch and tensor-network simulation architecture, we experimentally realize multi-wavelength optical tensor-matrix multiplication. In our previous work(*46*), we introduced an all-optical matrix-matrix multiplication scheme based on a single propagation of coherent light. Here, we extend this approach to 8 wavelength channels and changes the computational primitive from matrix-matrix to tensor-matrix multiplication within a single optical propagation, providing the higher-order structure subsequently used to organize operator and state-batch dimensions.

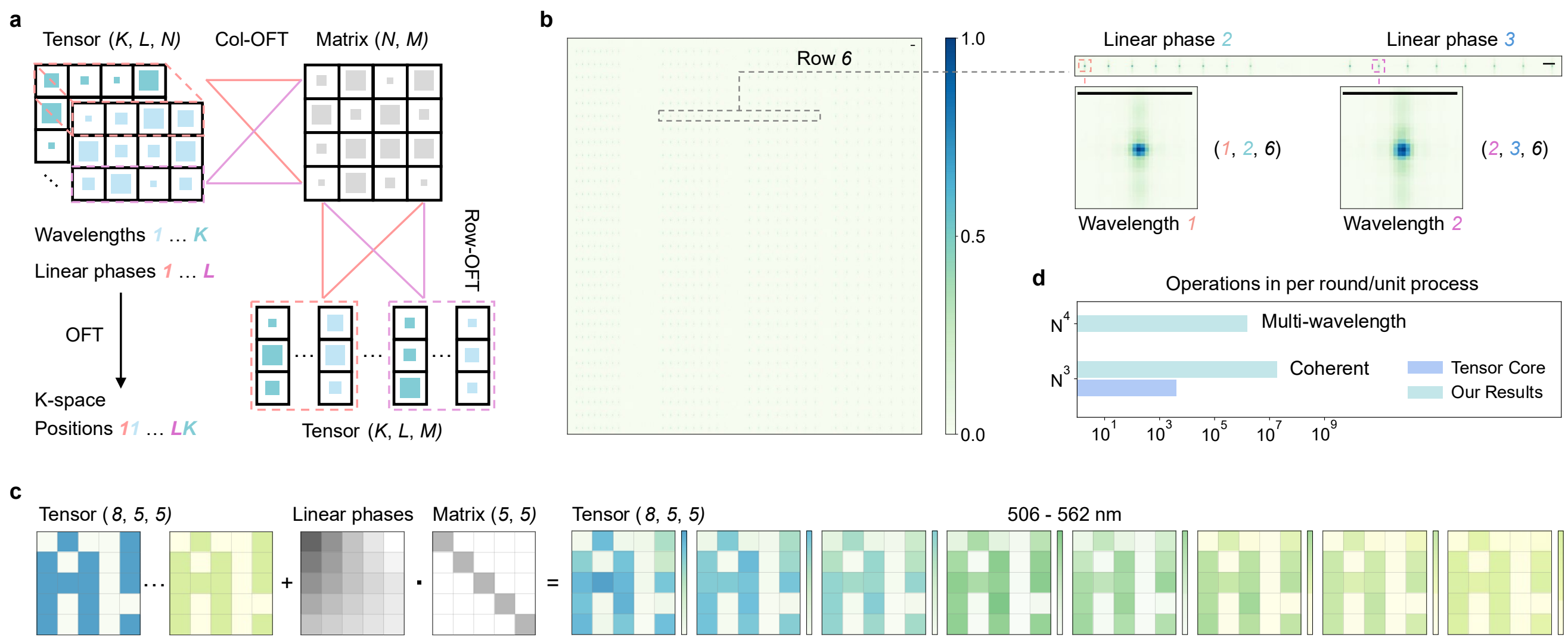

**Fig. 2 Demonstration and evaluation of parallel optical tensor-matrix multiplication. a** Principle of parallel optical tensor–matrix multiplication. OFT, optical Fourier transform. Different colors denote different wavelength channels and the linear phase gradients used for row encoding. **b** Numerical simulated optical field for example. A tensor-matrix multiplication of dimensions $(8, 4, 3000) \cdot (3000, 30) = (8, 4, 30)$ is shown, with all elements set to 1 and the output intensity displayed after max normalization. The scale bar corresponds to 100 µm. The resulting third-order tensor can be directly read out from the amplitude distribution in the two-dimensional spatial domain. **c** Experimental demonstration of parallel optical tensor–matrix multiplication. The experimental setup is shown in Fig. S1c, and the corresponding raw optical fields are presented in Fig. S2b. Except for the schematic illustration of the linear phase gradients, all other input and output patterns correspond to experimentally measured results and are directly retrieved from the raw optical fields. **d** Comparison between a single GPU tensor core and optical tensor processing in terms of operation granularity and parallelism. The horizontal axis indicates the number of operations represented in a single processing round or unit operation, and the vertical axis indicates the degree of parallelism. The comparison is based on a single GPU tensor core and the present experimental demonstration (details in Table S1).

As illustrated in Fig. 2a, a third-order tensor of dimensions $(K, L, N)$ is encoded jointly in the two-dimensional spatial amplitude distribution, which represents the $(L, N)$ dimensions, and in $(K)$ discrete wavelength channels. Within each wavelength channel, the $(L)$ rows of the spatial matrix are labelled using distinct linear phase gradients. During optical propagation, each wavelength-resolved tensor slice is multiplied by a spatially encoded matrix of dimensions $(N, M)$. The row-dependent phase gradients map the corresponding outputs to distinct positions in the spatial-frequency domain, while the wavelength-dependent spatial displacement produced by the optical Fourier transformation further separates the ($K$) spectral channels. The resulting ($K$, $L$) output groups can therefore be directly resolved in the k-space plane (detailed in Supplementary Note 1). This joint spatial and wavelength multiplexing enables the complete tensor-matrix multiplication to be performed in a single optical propagation, without requiring wavelength-resolved or other specialized detection (Fig. 2b). In the experiment, eight discrete wavelengths ranging from 506 to 562 nm, with an 8 nm spacing, are injected simultaneously. A test pattern, "A!", is encoded into the two-dimensional optical amplitude and replicated across the 8 wavelength channels, after which all wavelength-resolved inputs are simultaneously multiplied by a diagonal matrix (Fig. 2c). The measured outputs show good agreement with the corresponding references. Additional experiments using different spatial patterns and extended tests across both the spatial and wavelength dimensions further verify the reliability of the proposed method (details in Figs. S2 and S3).

The experimental results establish the feasibility of high-order tensor operations through spatial-spectral optical multiplexing. Under idealized scaling and individual control of each wavelength, a single optical propagation provides $N^4$ parallelism, corresponding to a task requiring $O(N^4)$ digital arithmetic operations, compared with the $N^3$ operation granularity of GPU tensor core matrix multiplication. For the analysed dimensions, this increases the arithmetic workload per physical operation from approximately $10^3$ to $10^6$. High-speed modulators and parallel detectors could translate this granularity into high effective throughput; however, end-to-end performance also depends on encoding and detection bandwidths, signal-to-noise ratio, calibration and data-transfer overheads. The comparison therefore describes operation-level parallelism rather than a measured wall-clock speedup over a state-of-the-art GPU.

## Classical simulations of quantum systems with high-order optical tensor processing

To demonstrate quantum circuit simulation using the parallelism of high-order optical tensor processing, we consider two computationally demanding tasks in the digital simulation of quantum systems: batch gate-state simulation (Fig. 3) and tensor network contraction (Fig. 4). Batch gate-state simulation involves applying multiple quantum gates to multiple quantum states, or equivalently to the columns of a density matrix, and can therefore be formulated as a tensor-matrix multiplication (Fig. 3a). As a proof of concept, we experimentally implement three distinct 4-qubit gates acting on all 16 computational-basis states using different wavelength channels (Fig.

3b, c). Specifically, the matrix representations of a multi-controlled-X gate at 500 nm, a controlled-SWAP gate at 520 nm, and a multi-controlled-Z gate encoded using separate wavelength channels for its positive and negative elements at 540 and 580 nm, respectively, are arranged into a tensor of dimensions (*4*, *16*, *16*). This tensor is then applied to a (*16*, *16*) matrix whose columns comprise all 4-qubit computational-basis states, from $|0000\rangle$ to $|1111\rangle$. Since the results in Fig. 2 have already demonstrated the parallel processing and natural separation of different wavelength components, the measurements at the different wavelengths were performed separately in the present proof-of-concept experiments, as described in the Methods. We further evaluate the method using representative 2- and 3-qubit gates (Fig. 3d). The 2-qubit experiments include CNOT at 510 nm, SWAP at 520 nm, and CZ encoded at 530 and 540 nm, yielding average fidelities of ~0.921, ~0.870, and ~0.828, respectively (details in Fig. S4). The 3-qubit experiments include controlled-SWAP at 500 nm, Toffoli (CCNOT) at 520 nm, and CCZ encoded at 540 and 560 nm, with corresponding average fidelities of ~0.791, ~0.822, and ~0.732 (details in Fig. S5). The output states are obtained directly from the measured optical intensity distributions after row-wise min-max normalization and Euclidean normalization, without model fitting or post-selection. Although the directly retrieved 4-qubit results exhibit relatively low fidelities because of the limited experimental signal-to-noise ratio, subtracting independently measured dark-field backgrounds increases the fidelities to ~0.854, ~0.848, and ~0.857 for the three gates, respectively (details in Fig. S6). These experiments are not intended to implement a specific quantum application but instead establish the feasibility of using high-order optical tensor processing for classical batch simulation of multiple gate-state combinations.

An important advantage of batch simulation is that it can reduce the number of sequential evaluations required during quantum algorithm development and circuit optimization. In practical algorithm design, a circuit often needs to be evaluated for multiple function instances and input states to verify its logical behavior and identify possible implementation errors. Conventional digital processors generally perform these gate-state combinations through repeated simulation runs, whereas the proposed optical tensor processing allows the full batch to be evaluated within a single optical propagation. To illustrate this capability theoretically, we consider the Deutsch algorithm(*47*), a foundational quantum algorithm that distinguishes constant functions ($f(x) = 0$ or 1) from balanced functions ($f(x) = x$ or $1 - x$), using a single measurement in a 2-qubit circuit (Fig. 3e). Because the oracle ($U_f$) depends on the selected function, the complete set of four Deutsch circuits is naturally represented by a tensor of dimensions (*4*, *4*, *4*) tensor, rather than by a simple (*4*, *4*) matrix. Evaluating these circuits for all computational-basis inputs can therefore be formulated as a tensor-matrix multiplication. In our numerical optical propagation model, the complete set of gate-state combinations is evaluated in a single propagation using the proposed 8-wavelength optical tensor-processing architecture (510-540 nm for positive values and 550-580 nm for negative values, details in Fig. S7). The results reproduce the expected logical behaviour of the Deutsch algorithm and show the importance of initializing the ancillary qubit ($q_0$) in the $|1\rangle$ state. For $q_1 = |0\rangle$ and $q_1 = |1\rangle$, measurement of the query qubit yields $|0\rangle$ for constant functions and $|1\rangle$, for balanced functions. When $q_1 = |1\rangle$, the corresponding outcomes are reversed (Fig. 3f). A single optical propagation therefore provides the complete mapping between the function type, input configuration, and measured output. Across all simulated cases, the maximum, minimum, and average state fidelities are ~1.000, ~0.982, and ~0.994, respectively (Fig. 3g). Together, these results demonstrate that high-order optical tensor processing provides a parallel framework for evaluating multiple quantum circuits and input states, with the potential to reduce the number of sequential simulation rounds required for quantum algorithm development and optimisation.

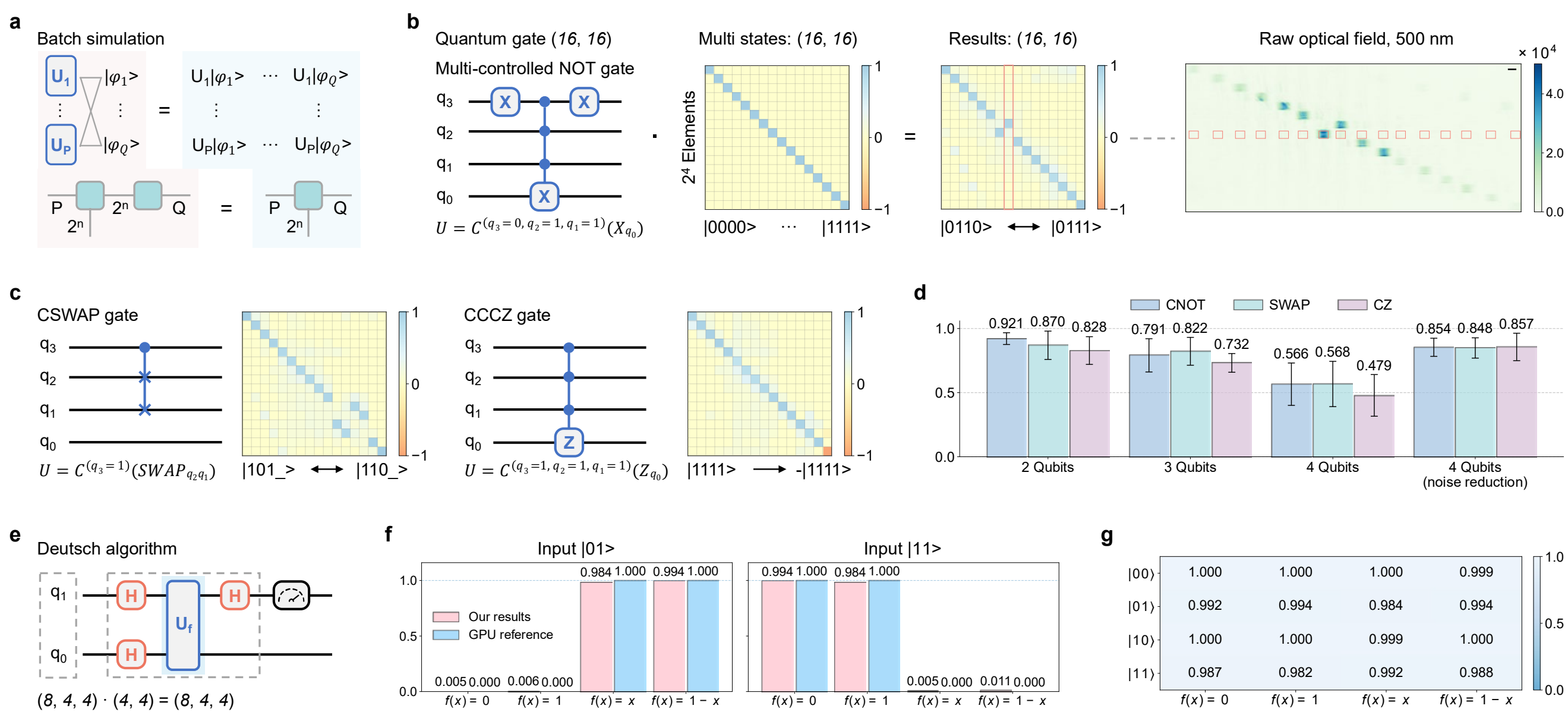

**Fig. 3 Parallel optical batch simulation of quantum gate and algorithm. a** Principle of batch gate-state simulation. In an n-qubit system, simulation of P quantum gates operation on Q quantum states can be formulated as a tensor–matrix multiplication between a tensor of dimensions ($P$, $2^n$, $2^n$) and a matrix of dimensions ($2^n$, $Q$). **b** Representative experimental results for a 4-qubit multi-controlled-X gate applied to all 16 computational-basis states. Left, inputs and retrieved output-state matrix (with background subtraction). Right, corresponding raw optical intensity distribution. The scale bar represents

20 detector pixels (92 μm). The red regions of interest (ROIs) along each row indicate the optical signals associated with one of the output states. The X, or NOT, gate exchanges $|0\rangle$ and $|1\rangle$. In the implemented multi-controlled-X operation, the states $|0110\rangle$ and $|0111\rangle$ are changed, while the remaining basis states are unchanged. **c** Experimental results for the controlled-SWAP (CSWAP) and multi-controlled-Z (CCCZ) gates (with background subtraction). The SWAP gate exchanges two qubits, whereas the Z gate introduces a phase shift of π to the $|1\rangle$ component. The CSWAP gate exchanges qubits $q_2$ and $q_1$ when the control condition is satisfied, whereas the CCCZ gate reverses the signs of the amplitudes associated with $|1110\rangle$ and $|1111\rangle$. **d** State fidelities obtained for representative 2-, 3- and 4-qubit gates. Each bar denotes the mean fidelity over all computational-basis input states: 4 states for 2-qubit gates, 8 states for 3-qubit gates and 16 states for 4-qubit gates. Error bars indicate ±1 standard deviation. As the experimentally retrieved amplitudes are real-valued, the fidelities are calculated as the squared inner products between the normalized measured and theoretical real-amplitude vectors. **e** Framework of the Deutsch algorithm. **f, g** Batch simulation of the Deutsch algorithm using numerical optical propagation. **f** Output probabilities obtained for the four possible oracle functions. Left, input state $|01\rangle$; right, input $|11\rangle$. **g** State-fidelity matrix for all combinations of computational-basis input states and oracle functions.

In addition to batch simulation, optical tensor processing can be applied to tensor network methods(*12*), which provide efficient approximate representations of quantum systems when entanglement remains sufficiently limited. The central idea is to decompose an exponentially large quantum-state vector into a network of lower-order tensors and to reproduce quantum evolution and measurement through a sequence of tensor contractions. Among the available tensor network representations, the MPS is one of the most widely used and can be efficiently implemented using GPU tensor kernels. In an MPS, a multi-qubit state is represented as a chain of local tensors, and a single-qubit gate is applied by contracting a gate matrix with the physical index of the corresponding tensor (Fig. 4a). This local operation is therefore directly compatible with optical tensor processing, in which the matrix-tensor contraction is performed through optical propagation. To demonstrate this correspondence, we numerically simulated a 7-qubit Grover algorithm(*48*) by replacing the single-site gate contractions in an otherwise digital MPS workflow with contractions calculated using the numerical optical tensor processing. Grover's algorithm reduces the number of oracle queries required for an unstructured search from $O(N)$ to $O(\sqrt{N})$, where $N = 2^n$ is the search-space dimension and n is the number of qubits (Fig. 4b). For the present 7-qubit implementation, the MPS bond dimension does not exceed $\chi = 2$. Each Grover iteration contains two layers of single-qubit Hadamard gates, corresponding to $7 \times 2$ local matrix-tensor contractions. Reaching the $9^{th}$ iteration therefore requires 126 such contractions. In the numerical optical implementation, each of these local contractions is replaced by optical propagation. The oracle and reflection operations, MPS canonicalization, and singular-value decompositions (SVD) remain digital; thus, the optical parallelism demonstrated here applies to the tensor elements within each local contraction rather than to the entire MPS workflow (details in Fig. S8).

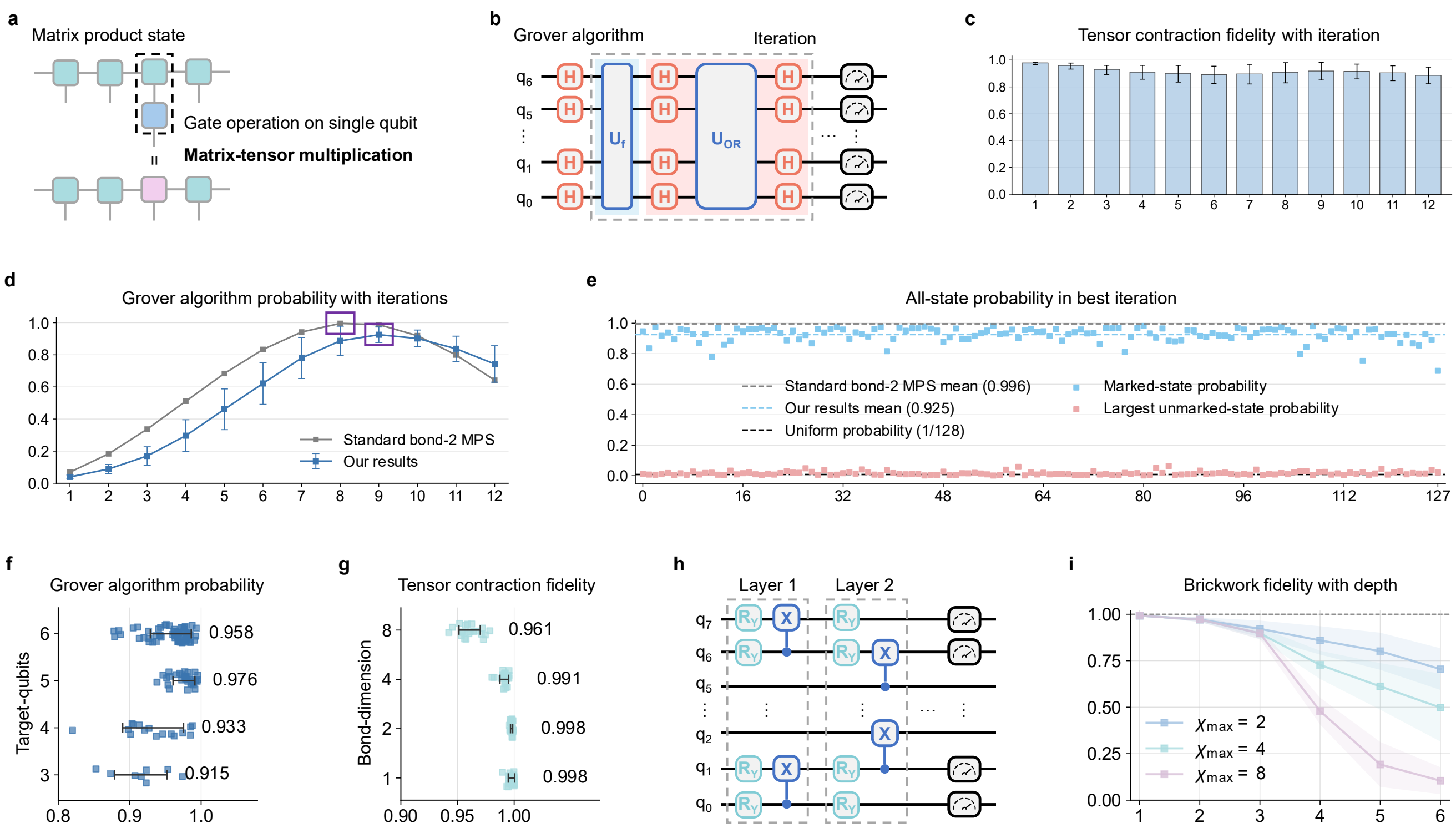

**Fig. 4 Parallel optical simulation of quantum tensor networks using a numerical physical-optics model. a** MPS representation of a multi-qubit state. A single-qubit gate acts on the physical index of the corresponding MPS tensor and is implemented as a matrix-tensor contraction. **b** Circuit framework of Grover's algorithm. The oracle ($U_f$) is shown in blue, whereas the diffusion operator, comprising Hadamard layers and the reflection ($U_{or}$), is shown in red. One oracle-diffusion cycle constitutes a Grover iteration. **c** Full-state fidelity between the numerically simulated optical MPS and the standard bond-2 MPS reference after each iteration. Bars and error bars represent the mean and ±1 s.d., respectively, over all 128 possible marked-state targets in the 7-qubit search space (details in Fig. S10a). **d** Mean marked-state probability as a function of Grover iteration for the numerical optical simulation and the standard bond-2 MPS reference. Error bars denote ±1 s.d. over the 128 target instances (details in Fig. S10b). **e** Marked-state probability and the largest unmarked-state probability for all 128 target instances. The standard bond-2 MPS results are evaluated at their optimal 8$^{th}$ iteration, whereas the numerical optical results are evaluated at their optimal 9$^{th}$ iteration. Dashed lines indicate the corresponding mean probabilities and the uniform probability (1/128). **f** Marked-state probabilities for 3-6-qubit Grover algorithms, evaluated at the standard Grover-optimal iteration for each system size. Each point represents one marked-state target, and horizontal bars denote the mean ±1 s.d. over all $2^n$ targets. **g** Fidelity of individual dense real-valued MPS tensor contractions at bond dimensions. Each point represents one independently generated random tensor, with 20 samples per bond dimension; horizontal bars denote the mean ±1 s.d. **h** 8-qubit brickwork circuit used for the depth benchmark. Each layer comprises independent single-qubit ($R_y$) rotations followed by alternating nearest-neighbor CNOT patterns. **i** End-to-end state fidelity between the closed-loop optical MPS

simulation and the exact state-vector reference as a function of brickwork depth for maximum bond dimensions. Symbols and shaded regions represent the mean and ±1 s.d., respectively, over 10 independently generated circuits with random $R_y$.

We first evaluated the full-state fidelity between the numerical optical MPS and the standard bond-2 MPS reference over 12 iterations (Fig. 4c). The target-averaged fidelity ranged from 0.885 to 0.975, with a mean of 0.914 across the 12 iterations. The accumulated contraction error caused the amplification trajectory to lag the standard Grover result, shifting the maximum mean marked-state probability from iteration 8 to iteration 9 (Fig. 4d). At these respective optimal iterations, the standard bond-2 MPS achieved a mean marked-state probability of 0.996, whereas the numerical optical simulation achieved 0.925 ± 0.048. Across the 128 target instances, the numerical optical probabilities ranged from 0.688 to 0.981. Nevertheless, the marked state remained the most probable output for every target, giving a top-1 success rate of 128/128 (Fig. 4e). The minimum probability margin between the marked state and the largest unmarked state was 0.666. We further evaluated bond-2 Grover searches containing 3-6 qubits using the same optical configuration and the standard Grover-optimal iteration for each system size (Fig. 4f). The mean marked-state probabilities were approximately 0.915, 0.933, 0.976, and 0.958 for 3, 4, 5, and 6 qubits, respectively. Together with the 7-qubit results, these measurements establish that the numerical optical processor can reproduce the amplification dynamics of Grover circuits when the MPS bond dimension remains small.

To quantify the dependence on bond dimension independently of circuit evolution, we next applied a Hadamard gate to dense, normalized real-valued MPS tensors of shape ($\chi$, 2, $\chi$), with $\chi$ = 1, 2, 4, and 8 (Fig. 4g). 20 independently generated tensors were evaluated at each bond dimension. The mean single-contraction fidelities were approximately 0.998, 0.998, 0.991, and 0.961, respectively. The reduction with increasing $\chi$ arises because larger tensors require more wavelength-spatial-phase output channels, thereby increasing the accumulated influence of channel overlap, diffraction sidelobes, and readout non-uniformity(*46*). Finally, we examined how these local errors accumulate with circuit depth using an 8-qubit real-valued brickwork circuit (Fig. 4h). We propagated the same random circuits using an exact state vector, a digitally truncated MPS, and a closed-loop optical MPS with $\chi_{max}$ = 2, 4, or 8 (details in Fig. S11). The $R_y$ contractions were calculated optically, whereas the CNOT updates and SVD truncations were performed digitally. As shown in Fig.4i, the end-to-end fidelity remained over 0.90 through depth 3 for all three bond limits but subsequently decreased with depth and bond dimensions. At depth 6, the mean fidelities were approximately 0.74, 0.50, and 0.10, respectively.

These numerical results establish the feasibility of optical tensor processing for shallow, low-entanglement tensor network simulations, while also identifying the principal limitation of the present architecture. Unlike many optical neural network tasks, iterative tensor network simulations require very small local errors to remain controlled over long contraction sequences. Consequently, a single-contraction fidelity of approximately 0.96 does not guarantee stable end-to-end evolution: coherent channel-dependent errors can accumulate and produce a substantially larger state-level deviation in deeper circuits (detailed error analysis in Supplementary Note 3). Scaling to larger bond dimensions and depths will therefore require reduced the crosstalk in wavelength and spatial-channel separation, more uniform readout and calibration, and tighter integration of optical contractions with digital canonicalization and truncation operations.

## Discussion

Previous classical optical simulations of quantum algorithms can exploit parallelism within a single gate-state transformation. Nevertheless, direct state-vector simulation requires a separately resolvable classical optical mode for each probability amplitude. The required spatial, spectral or temporal resources therefore still grow exponentially with qubit number, and optical parallelism alone does not remove this fundamental scaling. Our study establishes a classical optical architecture for parallel quantum-circuit simulation based on high-order optical tensor processing. For batch simulation, this architecture provides a higher degree of parallelism than conventional digital and optical methods. For large-scale simulation, the architecture naturally matches tensor network contractions. This work represents a research direction distinct from conventional optical neural networks (detailed comparison in Table S2).

We experimentally demonstrate 8-wavelength tensor-matrix multiplication within a single propagation and batched processing of complete computational-basis input sets for 2- to 4-qubit gates. Using a physics-based optical propagation model, we further extend the architecture to the Deutsch algorithm and to MPS simulations of the Grover algorithm and brickwork circuits. The architecture processes dynamically reconfigurable operators, state batches and virtual-bond contractions determined by the structure of quantum circuits. A practical implementation would therefore be hybrid: optical hardware would accelerate large and repetitive linear contractions, while electronic hardware would perform state storage, singular-value decomposition, truncation and adaptive control. The present results establish this computational organization rather than a measured end-to-end advantage over digital hardware. Continued advances in independently programmable wavelength channels, high-speed modulators, parallel detectors and integrated optoelectronic control could translate this architectural concurrency into practical throughput (details in Fig. S12). More broadly, this approach introduces quantum-simulation workloads into classical photonic computing and provides a potential bridge between advances in optical processors and the growing computational demands of quantum technologies.

## Materials and Methods

### Optical prototype

All experimental demonstrations of optical tensor processing (Fig. 2c) and batch simulation of multiple quantum gates (Fig. 3b-d) were performed using the optical prototype described here. In an ideal implementation (details in Fig. S1b), distinct spatial matrices would be independently encoded onto different wavelength channels to construct a complete third-order tensor, and all optical elements would be fully achromatic to ensure wavelength-independent propagation and modulation. In the present proof-of-concept system, component availability and fabrication constraints prevented the implementation of wavelength-selective spatial encoding. We therefore employed a hybrid eight-wavelength source consisting of a broadband supercontinuum laser and an eight-channel tunable optical filter, together with a single amplitude spatial light modulator (ASLM) for spatial-pattern encoding. Consequently, during a simultaneous tensor-matrix multiplication measurement, all wavelength channels carried the same spatially encoded pattern, and the tensor was replicated along the wavelength dimension.

This configuration nevertheless enabled simultaneous operation and readout across eight wavelength channels, thereby allowing us to verify the feasibility of spatial-spectral optical tensor processing. By contrast, the batch quantum-gate simulations required different gate matrices to be encoded at different wavelengths. Because the present system could not independently modulate the spatial pattern of each wavelength channel, measurements for different wavelength-matrix combinations were performed separately under otherwise identical experimental conditions and subsequently assembled to reconstruct the complete tensor output.

**Numerical optical simulation**

The batch simulation of the Deutsch algorithm (Fig. 3e-g), the tensor network-based simulation of the Grover algorithm (Fig. 4c-g), and the brickwork circuit (Fig. 4i) were performed using a numerical model of the optical tensor processor. Optical propagation was calculated using the standard angular-spectrum method(*49*), with free-space diffraction evaluated between each pair of successive optical elements. The third-order optical tensor was represented jointly by the wavelength dimension and a two-dimensional spatial grid. Each wavelength channel was propagated independently using its corresponding wavelength-dependent transfer function, while its complex optical field was retained throughout the propagation. At the detector plane, the wavelength-resolved complex fields were converted into intensity distributions. The detector was assumed to be insensitive to optical phase and to have an integration time substantially longer than the beat periods between different wavelength channels. Consequently, cross-wavelength interference terms were averaged out, and the detected intensity at each spatial position was calculated as the incoherent sum of the wavelength-resolved intensity contributions. We have open-sourced all the simulation code, including all details and configurations, and the details are in Supplementary Note 2. The simulation platform is configured with a NVIDIA® GeForce RTX™ 5090 GPU (32 GB dedicated memory), an Intel® Core™ Ultra 7 CPU, running Windows 11, Python 3.13.7, and PyTorch 2.8.0 with CUDA 12.9.

## Acknowledgements

**Funding**

This work is supported by the European Union's Horizon research and innovation program (101281131-QUEST, granted to Y.Z.), in part by the National Key Research and Development Program of China under contract nos. 2023YFB2804702 (granted to X.G.), in part by the Natural Science Foundation of China under contract nos. 62550072 (granted to X.G.) and nos. 62341508 (granted to Y.S.), in part by the Research Council of Finland Flagship Programme (320167, Aalto University), and in part by Shanghai Science and Technology Innovation Action Plan under contract nos. 25LN3201000 (granted to X.G.) and nos. 25JD1405500 (granted to X.G.). We gratefully acknowledge the Shanghai Institute of Optics and Fine Mechanics, Chinese Academy of Sciences, for providing access to an 8-wavelength tunable filter for our experiments.

We also express our appreciation to Alexandru Paler for his insightful comments on this paper. Z.S. thanks Profs. Jukka Pekkola and Pertti Hakonen and Dr. Minna Günes for bringing his group into the Centre of Excellence in Quantum Technology. This paper is dedicated to the Centre. This work is part of the Finnish Centre of Excellence in Quantum Materials (QMAT).

**Author Contributions**

Y.Z., X.G., and Z.S initiated the project. Y.S., K.W., X.G., and Z.S. supervised the project. Y.Z. and Z.S. conceived the idea and designed the experiments. Y.Z. performed the simulation and the experiments. All the authors analysed the results. Y.Z., Z.S., and X.G. prepared the manuscript. Y.Z., Z.S., X.G., Z.N., Z.Y., and S.T. revised the manuscript.

**Competing interests**

The authors declare that they have no competing interests.

## Data, code, and materials availability:

All raw experimental data are available in the main text and the supplementary materials, as well as via figshare at: https://doi.org/10.6084/m9.figshare.33191616. The code and results for numerical optical quantum simulation are available via GitHub at https://github.com/EkkoZYF/optical-tensor-quantum-simulation.